\documentclass[12pt,a4paper]{article}
\usepackage{graphicx,amsmath,amsfonts}

\def\,{\ifmmode\mskip\thinmuskip\else\leavevmode\thinspace\fi}

\newcommand{\nn}{\nonumber}
\newcommand{\vecc}[1]{\mathbf{#1}}
\newcommand{\veck}[1]{\mathbf{#1}{\lower-.2em\hbox{}^{2}}}
\newcommand{\vecs}[1]{\mathbf{#1}{\kern-.2em\hbox{}^{2}}}
\def\bs{\boldsymbol}

\begin{document}

\title{Multiple lepton pair production in relativistic ion collisions}
\author {E.~Barto\v s$^{1,2}$,
S.~R.~Gevorkyan$^{1,3}$, E.~A.~Kuraev$^1$\\ and
N.~N.~Nikolaev$^4$}
\date{} % optional
\maketitle

\begin{center}
{
$^{1}$ \it Joint Institute of Nuclear Research, 141980 Dubna, Russia\\
$^{2}$ \it Dept. of Theor. Physics, Comenius Univ., 84248
Bratislava, Slovakia \\
$^{3}$ \it  Yerevan Physics Institute, 375036 Yerevan, Armenia\\
$^{4}$ \it Institut f. Kernphysik, Forshchungszentrum J\"ulich,
D-52425 J\"ulich, Germany and L.~D.~Landau Institute for
Theoretical Physics, \\142432 Chernogolovka, Russia\\
}
\end{center}

\begin{abstract}
We apply the Sudakov technique to description of the multiple
production of lepton pairs in peripheral collisions of
ultrarelativistic heavy ions. For heavy ions with $Z_1\alpha\sim
Z_2\alpha\ll 1$ one needs a careful treatment of the multiple
Coulomb exchange between colliding ions and screening effects,
whereas interaction of real or virtual lepton pairs with colliding
ions can be neglected. We demonstrate that while the inclusive
spectra are modified by multiple Coulomb exchange between the
colliding ions, the Coulomb corrections to the momentum integrated
multiplicity distributions do vanish. After transformation to the
impact parameter representation the probability of $n$ lepton pair
production is shown to obey the Poisson distribution. The relevant
cross section is obtained.
\end{abstract}

%\vspace*{2cm}
\section{Introduction}
The multiplicity and the distribution of lepton pairs produced in
the Coulomb fields of two colliding relativistic heavy ions are
closely connected to the problem of unitarity. When heavy ions
collide at relativistic velocities their Lorentz contracted
electromagnetic fields are sufficiently intense to produce a large
numbers of such pairs. Usually the process of lepton pairs
production is considered as pair creation in the classic Coulomb
potential of a charge moving along a straight line. Such an
approach allows one \cite{HTB} to investigate the impact parameter
dependent total probability of the pair creation  $P(b)$, which by
definition is connected with the total cross section $\sigma=\int
{P(b)d^2b}$. As was noticed in \cite{BB} the probability of single
pair production calculated to lowest order in fine structure
constant  $\alpha=\frac{e^2}{4\pi}=\frac{1}{137}$ (throughout we
use units for which $ c=\hbar=1$) at small impact parameters
exceeds one, thus violating unitarity. This excess begins at
impact parameters smaller than the Compton wavelength of the
electron $ \lambda_c=\frac{1}{m}=386\,$fm and at energies of
practical interest (RHIC \& LHC).

Allowance for the finite size of colliding nuclei doesn't remedy the
situation, because  that would affect only the impact parameters
comparable to t0he nuclei radii, which are much smaller than the
Compton wavelength  of the electron $ R\ll
\lambda_c $.

As was shown in \cite{B} this problem can be solved by taking into
account the possibility of multiple pair production, whose
relative contribution grows with energy and dominates  at small
impact parameters. Since this early publication there has been
much work has in this direction (see e.g.~\cite{LMS} and references
therein) with the common for all statement: the probability to
produce n lepton pairs in the Coulomb field of heavy ions
colliding at fixed impact parameter $b$ can be approximately
represented as a Poisson distribution i.e.
$P(b,n)=\frac{W^n(b)}{n!} e^{-W(b)}$, where W(b) is the average
multiplicity of pairs at a fixed impact parameter.

Because of the somewhat controversial situation in the subject and
its importance for the operation of relativistic heavy ion
colliders (RHIC \& LHC), in the present communication we revisit
the multiple pair production based on the powerful Sudakov
technique of calculation of high energy Feynman diagrams. Recently
it has been applied \cite{BGKN} to the calculation of Coulomb
corrections to single lepton pair production in relativistic heavy
ion collisions. Here we extend the Sudakov approach to get the
probability of multiple pair ($ n\ge 2$) production in
relativistic heavy ions collisions. For heavy ions with charge the
numbers satisfy the conditions $Z_1\alpha\sim Z_2\alpha\ll 1$,
$Z_1Z_2\alpha \gg 1$, one needs full allowance of multiple Coulomb
interaction of colliding nuclei, whereas the secondary interaction
of produced pairs (real or virtual) with the Coulomb fields of
colliding ions can be neglected.

The paper is organized as follows. In section~\ref{sec:2} we
introduce the necessary definitions and  discussed briefly the
Sudakov'derivation of the Born amplitude for single pair creation.
Then We show how one can sum the perturbative series to get the
amplitude  for the process of $n$ pairs production in a compact
analytical form.

In Section~\ref{sec:3} we consider the Coulomb interaction between
the colliding ions and show how it affects the amplitude and cross
section of the process under consideration. Apart from the
conventional multiple photon exchange between ions, this
interaction include also the so called "screening corrections"
which are the result of the ions interaction through the virtual
lepton pairs and are responsible for the probability of
vacuum--to--vacuum transition. In the language of the
Abramovski--Gribov--Kancheli unitarity rules \cite{AGK}, the
diagrams with lepton pair loops can be associated with the QED
model for the pomeron, and the n-pair production amplitudes can be
associated with the unitarity cuts through n exchanged pomerons.
In section~\ref{sec:4} we report the multiple production
amplitudes in the impact representation in which the summation of
the perturbation series for the n-pair production can be done in a
compact closed form. Finally, using this amplitude we obtain the
probability of $n$ pair production which can be cast in the
Poisson form, and the total cross section of all pair production
process.

\section{The amplitude of the $n$ pairs production process.}
\label{sec:2}

\begin{figure}[t] \centering
\includegraphics[scale=.8]{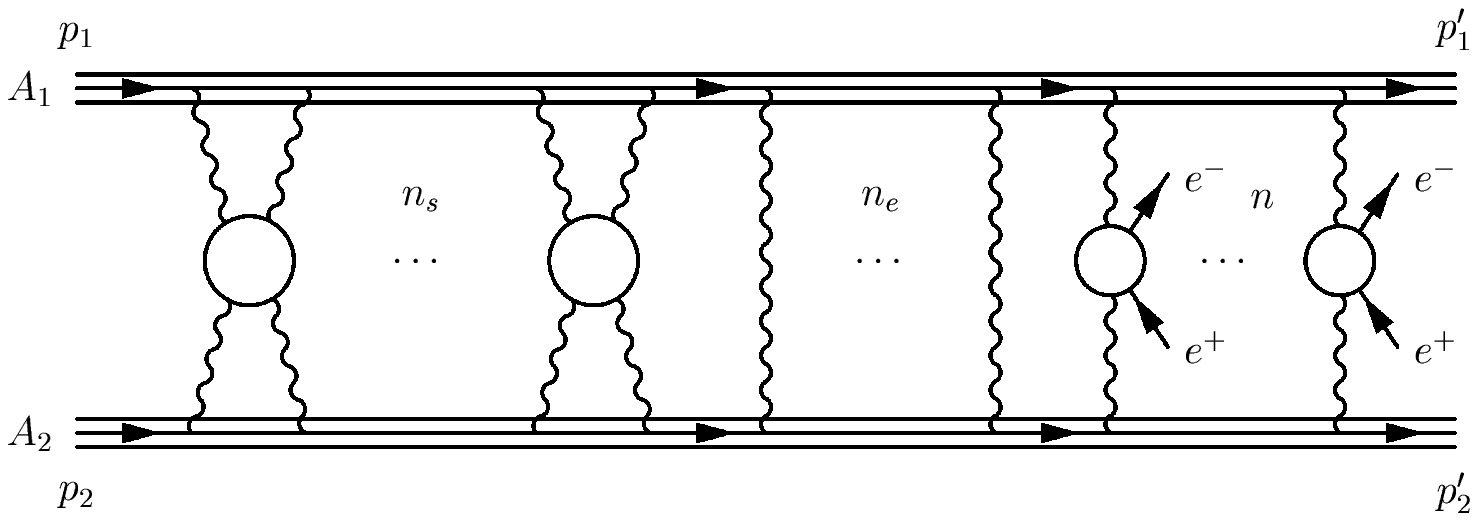}\vspace{.5cm}
\includegraphics[scale=.7]{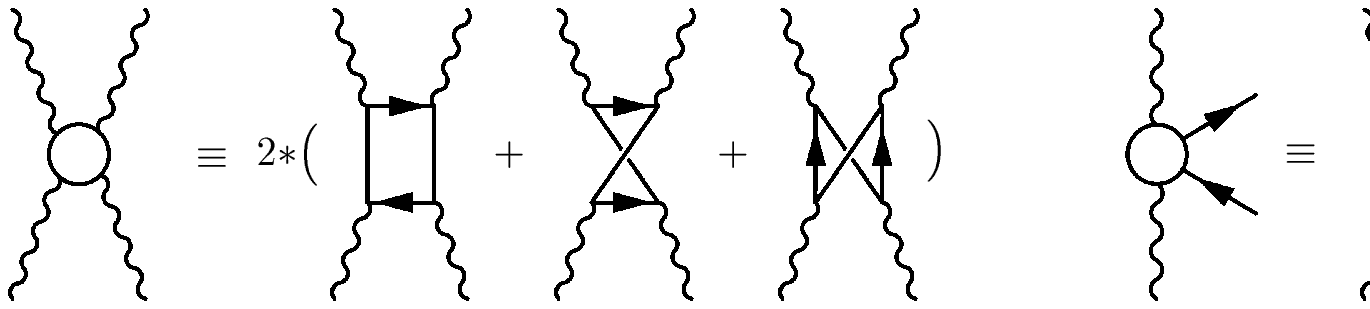}
\caption{The peripheral process of creation of $n$ lepton pairs
with  $n_e$ photon exchanges between nuclei $A_1$, $A_2$ and $n_s$
light--by--light scattering blocks.} \label{fig:1}
\end{figure}

The typical Feynman diagram (FD) describing the $n$ lepton  pairs
production in the collision of relativistic nuclei with atomic
numbers $A_1,A_2$, with $n_e$ exchanged photons between colliding
nuclei as well as screening effects e.g. the insertions of $n_s$
light by light (LBL) scattering blocks is drawn in
Fig.~\ref{fig:1}.

Upper and lower blocks in Fig.~\ref{fig:1} describes many virtual
photons interaction amplitudes with nuclei. They contain the
complete set of $(n_e+n+2n_s)!$ Feynman diagrams. To avoid the
multiple counting in what follows, we will multiply the relevant
amplitude by the factor $1/(n_e!n_s!(2!)^{2n_s})$.

As was mentioned above we restrict ourselves to ions with charge
numbers such that
\begin{equation}\label{}
Z_{1,2} \gg 1,\quad Z_1\alpha\sim Z_2\alpha\ll 1,\quad Z_1Z_2\alpha>1,
\end{equation}
which permits us to omit the multiphoton exchanges between the
produced pairs and colliding ions. The dominant mechanism is a
production of a single lepton pair per collision of equivalent
photons, i.e., one lepton loop per two-photon ladder. The
alternative mechanism of multiple pair production per collision of
equivalent photons, i.e., multiple lepton loops per two-photon
ladder, is suppressed by inverse powers of $Z_{1}Z_{2}$.

For the description of a peripheral process of n lepton pairs
creation i.e. the process
\begin{multline}\label{}
A_1(Z_1, p_1)+A_2(Z_2,p_2)\rightarrow A_1(Z_1,p_1')+
A_2(Z_2,p_2') + e_+e_-(r_1)+\dots + e_+e_-(r_n),\\ \quad
r_i=q_+^i+q_-^i
\end{multline}
It is convenient to use the Sudakov parameterization for the
four-momentum of all exchanged photons (for details see
\cite{BGKN})
\begin{gather} \label{}
k_i=\alpha_i\tilde p_2+\beta_i\tilde p_1+k_{i\bot},\quad
d^4k_i=\frac{s}{2} d\alpha_id\beta_id^2k_{i\bot},
\\ \nn s=(p_1+p_2)^2,\quad s\gg p_i^2=M_i^2\gg m^2,\quad
 \tilde p_1=p_1-p_2\frac{p_1^2}{s},\quad \tilde
p_2=p_2-p_1\frac{p_2^2}{s},\\ \nn \tilde p_1^2=\tilde
p_2^2=O\left(\frac{m^6}{s^2}\right),\quad \tilde p_1
k_{i\bot}=\tilde p_2 k_{i\bot}=0,\quad s= 2p_1p_2= 2\tilde
p_1\tilde p_2. \nn
\end{gather}
Here $\tilde p_i$ are light-like four-vectors build from $p_i$,
$M_i$ are the masses of colliding nuclei, $m$ and $s$ are the
electron mass and the total center mass energy.

The denominators of intermediate states of the nucleon Green
functions for upper and lower blocks are the same and have the
following form
\begin{equation}\label{}
s\sum_{i}\alpha_i- \Big(\sum_{i} \vecc{k_i}\Big)^2+i 0,\quad
s\sum_{i}\beta_i- \Big(\sum_{i} \vecc{k_i}\Big)^2+i 0.
\end{equation}

Peripheral process is characterized by small values of
longitudinal Sudakov parameters $\alpha_i$, $\beta_i$ and the
transverse momenta of the order of electron mass
\begin{equation}\label{}
|\alpha_i|\sim|\beta_i|\ll 1,\quad -k_{i\bot}^2\sim m^2.
\end{equation}
Further simplification follows from the form of the nominators of
the exchanged photons Green function (we work in the Feynman
gauge). Using the Gribov's representation for the metric tensors
\begin{equation}\label{}
g_{\mu\nu}=g_{\mu\nu}^{\bot}+\frac{2}{s}\Big(\tilde{p_1}^\mu
\tilde{p_2}^\nu+\tilde{p_1}^\nu \tilde{p_2}^\mu\Big).
\end{equation}
it is easy to show that for the typical conversion of the nuclei
currents $J_\mu(p_1)J^\mu(p_2)$ only one term, which contains the
scalar products of a nucleus current with a four--momentum of
another nucleus, becomes relevant (with power accuracy)
\begin{equation}\label{}
J_\mu(p_1) J^\mu(p_2)\approx \frac{2}{s}J_\lambda(p_1)p_2^\lambda
J_\sigma(p_2)p_1^\sigma\left(1+O\Big(\frac{m^2}{s}\Big)\right).
\end{equation}
It can be seen that the quantity $J_\mu(p_1) J^\mu(p_2)/s$ remains
finite with the large values of $s$. This fact provides great
simplification of the spinor structure of the amplitude
\begin{gather}\label{}
\bar{u}(p_1')\tilde p_2(p_1+\chi_1+M_1)\tilde p_2\dots(p_2+\chi_N+M_1)
\tilde p_2 u(p_1)\approx s^{N+1}N_1, \\ \nn
\bar{u}(p_2')\tilde p_1(p_2+\eta_1+M_2)\tilde p_1\dots(p_2+\eta_N+M_2)
\tilde p_1 u(p_2)\approx s^{N+1}N_2, \\ \nn
N_1=\frac{1}{s}\bar{u}(p_1')\hat{p}_2u(p_1),\quad
N_2=\frac{1}{s}\bar{u}(p_2')\hat{p}_1u(p_2).
\end{gather}
Besides we have $\sum|N_1|^2=\sum|N_2|^2=2$ for the nuclei with
the spin  1/2 and $|N_1|^2=|N_2|^2=1$ for the scalar one. Using
the identity
\begin{equation}\label{}
\sum_{perm}\frac{1}{\alpha_{i_1}}\frac{1}{\alpha_{i_1}+\alpha_{i_2}}
\cdots\frac{1}{\displaystyle\sum_{j=1}^N\alpha_{i_j}}=
\prod_{i=1}^N\frac{1}{\alpha_i}
\end{equation}
one can be convinced that the amplitude describing the upper and
lower blocks in Fig.~\ref{fig:1} can be put in the form
\begin{align}\label{}
I_1=N_1\prod_{i=1}^N\left(\frac{s}{-s\alpha_i+i0}+\frac{s}{s\alpha_i
+i0}\right),\;
I_2=N_2\prod_{i=1}^N\left(\frac{s}{-s\beta_i+i0}+\frac{s}{s\beta_i
+i0}\right)
\end{align}
with $N=n_e+2n_s+n-1 $.

This expressions contain all dependence on Sudakov parameters
$\alpha_i$, $\beta_i$ (the 4-momenta of exchanged photons in the
peripheral kinematics in denominators of their Green functions can
be considered as Euclidean two-vectors
$k_i^2=s\alpha_i\beta_i+k_{i\bot}^2\approx
k_{i\bot}^2=-\vecc{k_i}^2$).

At this stage the integration over Sudakov parameters can be done,
because the dependence of the amplitude on $\alpha_i$, $\beta_i$
provides the convergence of the relevant integrals
\begin{equation}\label{eq:11}
\int I_1\prod_{i=1}^Nd\alpha_i=(2\pi i)^N N_1,\quad \int
I_2\prod_{i=1}^Nd\beta_i=(2\pi i)^N N_2.
\end{equation}

Let us now consider the single pair production. The amplitude of
the process (1) in its lowest order (Born approximation) reads
\cite{BGKN}
\begin{equation}\label{eq:12}
M_{(0)}^{(1)}=is(8\pi\alpha)^2Z_1Z_2N_1N_2\frac{B_{\alpha\beta}
p_1^\alpha p_2^\beta}{s\vecs{q_1}\vecs{q_2}},\quad
B_{\alpha\beta}=\bar{v}(q_+)O_{\alpha\beta}u(q_-),
\end{equation}
$B_{\alpha\beta}$ is the Compton tensor \cite{BGKN} for pair
creation by two virtual photons with polarization vectors
$e_1(q_1)$, $e_2(q_2)$. $q_{1(2)}$ are the 4--momenta of exchange
photons and $r=q_++q_-$. Strictly speaking the squares of these
4--vectors $q_i^2$ do not vanish in the limit $\vecs{q_i}\to 0$.
This fact becomes essential when one calculates the total cross
section of a single pair production process. For the case of two
or more pairs production (which is our case) the replacement
$q_i^2=-\vecc{q_i}^2$ can  safely be done.

Using the  gauge invariance
\begin{equation}\label{eq:13}
q_1^\alpha B_{\alpha\beta}=q_2^\beta B_{\alpha\beta}=0
\end{equation}
one can perform the replacement
\begin{equation}\label{eq:14}
\frac{B_{\alpha\beta}p_1^\alpha p_2^\beta}{s}=\frac{B_{\alpha
\beta}e_1^\alpha e_2^\beta}{\tilde s_1}|\vecc{q_1}|
|\vecc{q_2}|,\quad
e_i^\alpha=\frac{q_{i\bot}^\alpha}{|\vecc{q_i}|}, \quad \tilde
s_1=s\alpha_2\beta_1,
\end{equation}
The quantity $\tilde s_1$ is related to the square of the
invariant mass of a pair
\begin{equation}\label{eq:15}
s_1=(q_1+q_2)^2=\tilde s_1-(\vecc{q_1}+\vecc{q_2})^2=
({q_+}+{q_-})^2.
\end{equation}
Two-dimensional vectors $e_i$ can be interpreted as a
polarization vectors of exchanged virtual photons.

Using (\ref{eq:13}-\ref{eq:15}) one can rewrite the Born amplitude
(\ref{eq:12}) in a form
\begin{subequations}
\begin{gather}
M_{(0)}^{(1)}=isN_1N_2B(q_1,q_2),\label{eq:16a}\\  B(q_1,q_2)=
(8\pi\alpha)^2 Z_1Z_2\frac{B_{\alpha\beta}e_1^\alpha e_2^\beta}
{\tilde s_1 |\vecc{ q_1}||\vecc{ q_2}|} \label{eq:16b}
\end{gather}
\end{subequations}
Now we are able to construct the amplitude for the process of $n$
pairs production. Bearing in mind the expressions (\ref{eq:11})
the matrix element of two pairs production  can be represented as
the convolution of two Born terms from expression (\ref{eq:16b})
\begin{equation}\label{}
M_{(0)}^{(2)}=i^2sN_1N_2\int B(k,k-r_1) B(q-k,q-r_2-k)
\frac{d^2\vecc{k}}{8\pi^2}.
\end{equation}
A straightforward generalization to the matrix element in the case
of $n$ pairs production reads
\begin{multline}\label{}
M_{(0)}^{(n)}=i^nsN_1N_2\int\prod_{i=1}^{n-1}
\left(B(k_i,k_i-r_i)\frac{d^2k_i}{8\pi^2}\right)B(h,h-r_n),\\
h=q-\sum_{i=1}^{n-1}k_i.
\end{multline}
Thus one can see that the amplitude for multiple pair production
is solely determined by the convolution of the amplitudes
corresponding to single pair production.

In the language of the AGK unitarity rules, this result
can be interpreted as unitarity cut through all exchanged
pomerons.

\section{The Coulomb exchanges between ions} \label{sec:3}

Let us now consider the effect of $m$ photon exchanges  between
nuclei $A_1$, $A_2$. The arguments given above leads to the
following  matrix element for the process of $n$ pairs production
with m photons exchanges among the colliding ions
\begin{align}\label{}
M_{(m)}^{(n)}&=\frac{i^nsN_1N_2}{m!}\int\prod_{j=1}^{m}
\left(\frac{-i \alpha
Z_1Z_2}{\bs{\chi}_j^2+\lambda^2}\frac{d^2\bs\chi_j}{\pi}\right)
\\ \nn
&\;\cdot\int\prod_{i=1}^{n-1}\left(B(k_i,k_i-r_i)\frac{d^2
\vecc{k_i}}{8\pi^2}\right)B(k_n,k_n-r_n),\;
k_n=q-\sum_{i=1}^{n-1} k_i -\sum_{i=1}^{m}\chi_i .
\end{align}

Another effect which we take into account is the possibility of
the ion--ion interaction through the LBL blocks (screening effect
in Fig.~1), which in the AGK language \cite{AGK} is equivalent to
the exchange by additional un--cut pomerons. It is associated with
the iteration of a typical kernel
\begin{subequations}
\begin{gather}
L\int Y(\vecc{l})d^2\vecc{l},\label{eq:20a}\\
Y(\vecc{l})=\frac{(\alpha^2Z_1Z_2)^2} {32\pi^4}\int
\frac{P}{|\vecc{l_1}||\vecc{l_2}||\vecc{l_1} -\vecc{l}|
|\vecc{l_2}+\vecc{l}|}d^2\vecc{l_1} d^2\vecc{l_2} \frac
{d\tilde{s_1}} {\tilde{ s_1}^2},\\
P=\Pi^{\alpha\beta\gamma\delta}e_1^\alpha(l_1)e_2^\beta (l_1-l)
e_3^\gamma(l_2)e_4^\delta(l_2+l),\label{eq:20c}
\end{gather}
\end{subequations}
where we rearrange the "extra" phase volume of longitudinal
Sudakov parameters $\alpha_2$, $\beta_1$ in terms of invariant
mass square of LBL block and extract explicitly the boost degree
of freedom of LBL block
\begin{figure}[t] \centering
\includegraphics[scale=.9]{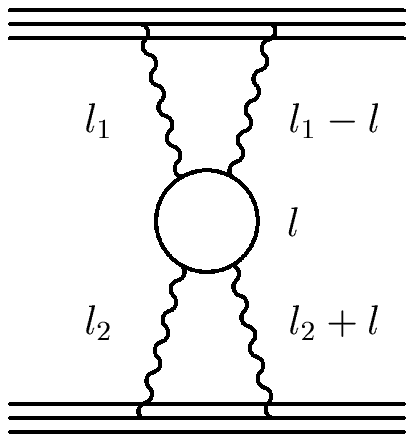}
\caption{Typical kernel of the ion--ion interaction through the
LBL blocks.} \label{fig:2}
\end{figure}
\begin{gather}\label{}
\int\frac{d\alpha_2 d\beta_1}{s(\alpha_2\beta_1)^2}=\int \frac{d
\beta_1}{\beta_1}\int\frac{d \tilde s_1}{\tilde s_1^2}=L\int
\frac{d\tilde s_1}{\tilde s_1^2},\\ \nn
L=\ln\gamma_1\gamma_2,\quad \tilde s_1=(l_1+l_2)^2>4m^2.
\end{gather}
The structure (20) entered the matrix element (19) in the compact
form
\begin{equation}
\frac{1}{n_s!}\prod_{i=1}^{n_s}\Big(LY(\vecc{l_i})d^2\vecc{l_i}\Big)
\end{equation}

Real part of (\ref{eq:20c}) (equal to one half of its s--channel
discontinuity) is related with (\ref{eq:16b})
\begin{multline}\label{}
\Re\Pi^{\alpha\beta\gamma\delta}e_1^\alpha(l_1)e_2^\beta (l_1-l)
e_3^\gamma(l_2)e_4^\delta(l_2+l)=\frac{1}{2}\int
B_{\alpha\beta}(l_1,r-l_1)e_1^\alpha(l_1) e_2^\beta(r-l_1)\\\cdot
B_{\gamma\delta}(l_1-l,r+l-l_1)e_1^\gamma(l_1-l)
e_2^\delta(r+l-l_1)d\Phi_r,
\end{multline}
where $d\Phi_r$ is the phase volume of the intermediate pair
\begin{equation}\label{}
d\Phi_r=\frac{\delta^4(r-q_+-q_-)d^3q_+d^3q_-}{(2\pi)^2
2\varepsilon_+2\varepsilon_-}.
\end{equation}
We will show later that the imaginary part of $P$ is irrelevant
either for the total cross section or for the probability of $n$
pairs production distribution.

\section{The impact parameter representation for the amplitude}
\label{sec:4}

The last step in building  the matrix element for the process  of
$n$ real pairs creation consist in  transformation of the obtained
above expressions in the impact--parameter representation and
summation over all eikonal photons and LBL blocks. For this we
introduce the identity
\begin{multline}\label{eq:25}
\int\delta^2\Big(\vecc{k_n}-\vecc{q}+\sum_{i=1}^{n-1}\vecc{k_i}+
\sum_{i=1}^{n_s}\vecc{\chi_i}+\sum_{i=1}^{n_v}\vecc{l_m}\Big)d^2
\vecc{k_n}= \\ \frac{1}{4}\int e^{-i\vecc{q}\bs\rho}
\exp\left[{i\bs\rho\Big(\vecc{k_n}+\sum_{i=1}^{n-1}
\vecc{k_i}+\sum_{i=1}^{n_s}\vecc{\chi_i}+\sum_{i=1}^{n_v}\vecc{l_m}
\Big)}\right]\frac{d^2\vecc{k_n}}{\pi}\frac{d^2\bs \rho}{\pi}=1\cdot
\end{multline}
Using this expression the summation in $n_e$ and $n_s$ can be
easily done with the result
\begin{equation}\label{eq:26}
M^{(n)}=\frac{i^n\pi s}{2}N_1N_2\int
e^{-i\vecc{q}\bs\rho}e^{i\Psi(
\rho\lambda)}e^{-L[A(\rho)/2+i\varphi(\rho)]}\prod_{i=1}^n
\tilde{B}(\bs\rho,\vecc{r_i})\frac{d^2\bs\rho}{\pi},\quad n\ge 2,
\end{equation}
with
\begin{gather}\label{eq:27}
\frac{A(\rho)}{2}+i\varphi(\rho)=\int Y(\vecc{l})e^{i\vecc{l}
\bs\rho}\frac{d^2\vecc{l}}{\pi},\quad \tilde{B}(\bs\rho,
\vecc{r_i})=\int{B(\vecc{k},\vecc{k}-\vecc{r_i})e^{i\vecc{k}\bs\rho}
\frac{d^2\vecc{k}}{8\pi^2}},\\
\Psi(\rho\lambda)=-\alpha Z_1Z_2\int\frac{e^{i\bs\chi\bs\rho}}
{\bs\chi^2+\lambda^2} \frac{d^2\vecc{\bs\chi}}{\pi}=-2\alpha
Z_1Z_2K_0(\lambda\rho)
\end{gather}
where $K_0(\lambda\rho)$ is modified Bessel function (Mac-Donald
function).

The phase volume of final state which consists from the scattered
nuclei and $n$ pairs can be written in the following form
\begin{align}\label{eq:29}
d\Gamma_{n+2}&=\prod_{i=1}^n\Bigg(\frac{d^3q_+d^3q_-} {(2\pi)^6
2\varepsilon_+2\varepsilon_-}\Bigg)\frac{1}{(2\pi)^2}
\frac{d^3p_1'd^3p_2'}{2\varepsilon_1'2\varepsilon_2'}\delta^4
\Big(p_1+p_2-p_1'-p_2'-\sum_{i=1}^nr_i\Big)\nn\\&=\prod_{i=1}^n
\Bigg(\frac{L}{(2\pi)^4}\frac{d^2\vecc{r_i}}{2}ds_id\Phi_i
\Bigg)\frac{d^2\vecc{q}}{2s(2\pi)^2}
\end{align}
with $q=p_{1\bot}'$.

Cross section of $n$ pairs production has the form
\begin{equation}
d\sigma_n=\frac{1}{8s}\frac{|M_n|^2}{n!}d\Gamma_{n+2}.
\end{equation}
Statistical factor $1/n!$ is included to take into account the
identity of pairs. Using the expressions
(\ref{eq:25})-(\ref{eq:29}) we get
\begin{equation}\label{eq:31}
\frac{d\sigma_n}{d^2\rho}= P_n(\rho),\quad
P_n(\rho)=\frac{(LA_1(\rho))^n}{n!}e^{-LA_1(\rho)},\quad n\ge 2
\end{equation}
with
\begin{equation}\label{eq:32}
A_1(\rho)=\frac{1}{2^5\pi^4}\int|B(\rho,r)|^2ds_1d^2\vecc{r}d\Phi_r.
\end{equation}
It can be easily recognized that for $A(\rho)$ from (\ref{eq:27})
\begin{equation}\label{eq:33}
A(\rho)=A_1(\rho),
\end{equation}
thus confirming the Poisson character of probability distribution
in impact-parameter representation.

Let us mention that the effect of eikonal photons as well as the
imaginary part of the amplitude corresponding to LBL blocks do not
modify the total cross section as well as differential cross
section integrated over phase volume of final nuclei. Really,
integrating the square of the amplitude (\ref{eq:26}) over the
phase volume one immediately obtains
\begin{multline} \label{eq:34}
\int e^{i\vecc{q}(\bs\rho_1-\bs\rho_2)}e^{i\{[\psi(\rho_1)-\psi
(\rho_2)] -L[\varphi(\rho_1)-\varphi(\rho_2)]\}}f(\rho_1)f(\rho_2)
\frac{d^2 \vecc{q}}{\pi}\frac{d^2\bs\rho_1}{\pi}
\frac{d^2\bs\rho_2}{\pi}=\\ \int
f^2(\rho)\frac{d^2\bs\rho}{\pi}\cdot
\end{multline}
Nevertheless, the exclusive cross section is sensitive to both
these factors.

Expression (\ref{eq:32}) can be simplified if one neglects the
dependence of Compton tensor (\ref{eq:14}) on external photons
virtualities
\begin{equation}\label{}
B_{\alpha\beta}(k,r)e^\alpha e^\beta\to
B_{\alpha\beta}(0,r)e^\alpha e^\beta
\end{equation}
and use the well known relation \cite{KL}
\begin{equation}\label{}
\int\limits_{4m^2}^{\infty}\sum\Big|B_{\alpha\beta}(0,r)e^\alpha
e^\beta\Big|^2\frac{ds_1d\Phi_r}{s_1^2}=\frac{7}{36\pi m^2}\cdot
\end{equation}
As a result the expression (\ref{eq:32}) can be cast in the form
\begin{equation}
A(\rho)=\frac{7}{18\pi^2m^2}(\alpha^2 Z_1Z_2)^2 I(\rho),
\end{equation}
\begin{align*}\label{}
I(\rho)&=\int\limits^m\frac{e^{i\bs\rho(\vecc{k_1}-\vecc{k_2})}}{|\vecc{k_1}|
|\vecc{k_1}-\vecc{r}||\vecc{k_2}||\vecc{k_2}-\vecc{r}|}
\frac{d^2\vecc{r}}{\pi}\frac{d^2\vecc{k_1}}{\pi}\frac{d^2\vecc{k_2}}
{\pi}\\&=\int\limits^m \frac{e^{i\bs\chi\bs\rho}}
{|\vecc{k}||\vecc{k}-\bs\chi||\vecc{k'}||\vecc{k'}+\bs\chi|}
\frac{d^2\bs\chi}{\pi}\frac{d^2\vecc{k}}{\pi}\frac{d^2\vecc{k'}}
{\pi}\\&=\int{e^{i\bs\chi\bs\rho}}\ln^2\left(\frac{m^2}
{\bs\chi^2}\right) \frac{d^2\bs\chi}{\pi} \approx
\frac{16}{\rho^2}\left(\ln(\rho m)+O(1)\right),
\end{align*}
where we introduce the cut--off parameter $|k|<m$ as a result of
the fast decreasing of matrix element of pair production by two
photons. For the case of heavy leptons production ($\mu$ or
$\tau$) the upper limit must be replaced by quantity $Q$ which
can be associated with maximal momentum transferred to nucleus
without its disintegration. For the case $\rho m\gg 1$ one has
\begin{equation}\label{}
A(\rho)\approx\frac{56}{9}\frac{(\alpha^2 Z_1Z_2)^2}{\pi^2(\rho
m)^2}\left(\ln(\rho m)+O(1)\right)
\end{equation}
which is in the agreement with \cite{LMS}.

Our formula (\ref{eq:31}) can be applied for the case $n=0$ (the
probability of elastic nuclei scattering). The case $n=1$ needs a
bit accurate consideration. The expression for $\sigma_1$ can be
cast in the following form
\begin{equation}\label{eq:39}
\sigma_1=\sigma_B+L\int A(\rho)\left(e^{-LA(\rho)}-1\right)
d^2\bs\rho.
\end{equation}
First term $\sigma_B$ corresponds  to the leading order of the
Racah formula (see for instance \cite{BB}) for the cross section
in Born approximation. Inferring (\ref{eq:39}), one has to take
into account the longitudinal components of momenta of exchanged
photons which create the pair. The second term takes into account
the unitarity corrections to the total cross section.

\section*{Conclusions}

In conclusions let us enumerate very shortly the main results of
the present work. In our paper we obtained the general form for
the amplitude of $n$ lepton pairs production, accounting for the
mutual Coulomb interaction of relativistic ions. We get the
probability of vacuum--vacuum transition in the closed analytic
form. We confirmed that the probability of $n$ pair production can
be approximated by Poisson distribution. Although this result has
been claimed before in a number of publications, the mutual
interaction of ions has not been taken into account in these
works. At last we have shown that the Coulomb interaction between
the colliding nuclei, although important in the exclusive
kinematics, does not affect the integral yield of lepton pairs.

\section*{Acknowledgements}

We are grateful to Valery Serbo, Alexander Tarasov and Lev Lipatov
for their interest to the subject and useful comments. The work is
supported by INTAS 97-30494, INTAS-00366 and SR-2000 grants. E.~K.
is grateful to NCPI, JINR for support.

\end{document}